\documentclass  {article}

\usepackage{amsfonts,amsmath,amscd}

\begin{document}

\title{%
Curved Momentum Space and Relative Locality  }
\author{J. Kowalski-Glikman\thanks{email: {\tt jkowalskiglikman@ift.uni.wroc.pl}}\\Institute for
Theoretical Physics,\\ University of Wroc\l{}aw,\\ Pl.\ Maksa Borna
9, Pl--50-204 Wroc\l{}aw, Poland} \maketitle

\begin{abstract}
I briefly discuss the construction of a theory of particles with
curved momentum space and its consequence, the principle of relative
locality.
\end{abstract}

\vspace{12pt}

\begin{center}
{\em Dedicated to Professor Jerzy Lukierski on his 75th
birthday.}\end{center}\vspace{16pt}

Today, almost one hundred years after  general relativity was first
formulated, we are accustomed to the notion of curved spacetime. We
know that the nontrivial spacetime geometry is in one to one
correspondence with the dynamical gravitational field and how to
relate the abstract mathematical structures of Riemannian (and, more
generally, Cartan) geometry with physical concepts like distance,
acceleration, tidal forces etc. Moreover, we understand that the
source of the gravitational field is the energy-momentum of matter
(and of gravity itself). This is possible because the dimension of
the product of Newton's constant and the energy density has the
dimension of inverse length, establishing in this way the curvature
scale ("matter tells geometry how to curve".)

Already at the dawn of differential geometry Carl Friedrich Gauss
noticed that the necessary prerequisite for the existence of a
nontrivial (curved) geometry is the presence of a scale:

\begin{quote}
``The assumption that the sum of the three angles [of a triangle] is
smaller than $180^\circ$ leads to a geometry which is quite
different from our (Euclidean) geometry, but which is in itself
completely consistent. I have satisfactorily constructed this
geometry for myself [\ldots], except for the determination of one
constant, which cannot be ascertained a priori. [\ldots]  Hence I
have sometimes in jest expressed the wish that euclidean geometry is
not true. For then we would have an absolute a priori unit of
measurement.''\footnote{As cited in \cite{Milnor}.}
\end{quote}

The necessity of the presence of the scale is easy to understand.
Indeed a nontrivial geometry requires nonlinear structures and those
can be constructed only if there is a scale that makes it possible
to construct nonlinear expressions from fundamental, dimensionfull
basic variables.

General relativity tells us that spacetime is curved, but what about
another spaces that are of relevance in physics? One can interpret
the Gauss' dictum as a statement that if a scale of some physical
quantity is present in a theory, one could expect that the geometry
of the corresponding manifold must be nontrivial. Or putting it in
other words: ``everything is curved unless it cannot be.''

There are several examples that support this claim. Special
relativity introduces a scale of velocity, and one could suspect
 that the manifold of (three) velocities may possess
nontrivial structures. And indeed it does! Contrary to Galilean
mechanics in special relativity the velocity composition law is
highly nontrivial
\begin{equation}\label{1}
    \vec{v}\oplus \vec{u} =
    \frac{1}{1+\vec{u}\vec{v}/c^2}\left(\vec{v}+\frac{\vec{u}}{\gamma_v}+\frac1{c^2}\frac{\gamma_v}{1+\gamma_v}\,
    (\vec{v}\vec{u})\vec{v}\right)\,, \quad
    \gamma_v=\sqrt{1-\vec{v}^2/c^2}\, .
\end{equation}
This expression is neither symmetric nor associative and is related
to deep mathematics \cite{Girelli:2004xy}. It also has interesting
physical consequences (Thomas precession).

The relativistic, four-momentum space is, arguably, even more
important physically than the spacetime. Indeed virtually all
physical measurements can be reduced to the measurements of energies
and momenta of incoming particles of various kinds (probes)
performed by measuring devices located at the origin of a coordinate
system. It is only by observing the incoming probes that we can
infer the properties of distance events
\cite{AmelinoCamelia:2011bm}, \cite{AmelinoCamelia:2011pe}. The
question arises as to if we have good reasons to believe that the
momentum space is an almost structureless Minkowski space, or it is
conceivable perhaps that it could posses more intricate geometrical
structures.

Following the Gauss' intuition a possible way of addressing this
question is to look for a theory that could provide us with a
momentum scale $\kappa$. Such a theory indeed exists, and is pretty
well known \cite{Witten:1988hc}.

In $2+1$ spacetime dimensions the Newton's constant $G_3$ has the
dimension of inverse mass raising the hope that it may provide the
sought  momentum scale being a prerequisite for the emergence of a
nontrivial momentum space geometry. This hope was fully confirmed by
the dynamical model calculations \cite{Matschull:1997du},
\cite{Meusburger:2003ta}.

Let us briefly recall the picture that emerges from these papers. As
it is well known, in $2+1$ dimensions gravity is described by a
topological field theory, so that  local degrees of freedom are not
present. Consider a point massive particle coupled to gravity. Since
the system {\em particle + gravity} has only a finite number of
degrees of freedom one can solve it exactly to obtain an effective
description of the particle that includes the back-influence of the
gravitational field it creates. As it turns out
\cite{Matschull:1997du} such effective theory can be described as a
theory of the particle with curved momentum space,  the curvature
scale being $G_3$, as expected. Similar conclusion has been reached
in the case of a scalar field coupled to quantum gravity in $2+1$
dimensions \cite{Freidel:2005me}: by integrating out in the path
integral (Ponzano--Regge model) the gravitational degrees of freedom
one obtains an effective action for the scalar field, which can be
interpreted as an action for the field with curved momentum space.

And what about gravity in the physical $3+1$ dimensions? Now the
Newton's constant is the ratio of  Planck length $\ell_P$ and the
Planck mass $M_P$
\begin{equation}\label{2}
    \ell_P=\sqrt{\hbar G}\,,\quad M_P=\sqrt{\frac\hbar G}\,,
\end{equation}
 and therefore has the dimension of length over mass.
However, one can imagine a regime, presumably of quantum gravity, in
which the Planck length is negligible, while the Planck mass remains
finite. This formally means that both $\hbar$ and $G$ go to zero, so
that both quantum and local gravitational effects become negligible,
while their ratio remains finite \cite{Girelli:2004md}. In more
physical terms this regime is realized if the characteristic length
scales relevant for the processes of interest are much larger than
$l_P$, while characteristic energies are comparable with the Planck
energy. An example of such a process might be the gravitational
scattering in the case when the longitudinal momenta are Planckian,
while the transferred momentum is very small (as compared to $M_P$)
\cite{'tHooft:1987rb}, \cite{Verlinde:1991iu}. In the case of such
processes we again encounter the situation that the momentum scale
is present, and we expect to find a nontrivial geometry of the
momentum space. Unfortunately, to date no specific  model of this
kind has been formulated.

In absence of concrete models derived directly from quantum gravity,
let us just assume that momentum space has a nontrivial geometry,
governed by the the momentum scale $\kappa$ (presumably of order of
the Planck mass) and let us try to derive the most general possible
description of particles kinematics. In order to do so it is
convenient to start with the discussion of the standard relativistic
particles action, to see how could it be generalized.

The free relativistic particle action
\begin{equation}\label{3}
    S=-\int_{-\infty}^{+\infty} d\tau\, x^a\dot p_a
    +N\left(\eta^{ab}p_ap_b-m^2\right)
\end{equation}
consists of two terms: the kinetic one $-x^a\dot p_a$,
$a,b=0,\ldots,3$  and the mass shell constraint
$\eta^{ab}p_ap_b-m^2$ imposed by the Lagrange multiplier $N(\tau)$.
It is worth noticing that the term $\eta^{ab}p_ap_b$ is nothing but
the square of the Minkowski distance between the point ${\cal P}$ in
momentum space, with coordinates $p_a$ and the momentum space origin
${\cal O}$ with coordinates $p_a=0$, calculated along the straight
line joining these two points, i.e., the geodesic of the Minkowski
space geometry.

The equations of motion resulting from this action are
\begin{equation}\label{3a}
    \dot p_a = 0\,, \quad \eta^{ab}p_ap_b=m^2\,, \quad \dot x^a = N\,
    \eta^{ab}p_b\,.
\end{equation}

It is also worth noticing that the action (\ref{3}) is manifestly
invariant under global Lorentz and local $\tau$ reparametrization
symmetries, as well as under the following global translations (up
to the boundary terms)
\begin{equation}\label{4}
   \delta x^a = \xi^a\,,\quad \delta p_a=\delta N =0\,.
\end{equation}

Finally, it should be stressed that the action (\ref{3}) can be
meaningfully written down only if the coordinates $x^a$ and $p_a$ on
both spacetime and momentum space are given {\em a priori}.

The model presented above was essentially the standard theory of
free relativistic particles. Let us now extend this theory to
describe a system of interacting particles. In order to do that we
make use of the intuitions coming from the theory of Feynman
diagrams. Let us describe the system with a single three-particles
interaction in details -- the generalization to the more complex
cases is going to be obvious.

Let us assume that we have to do with some number of incoming
particle and the outgoing ones, interacting in a vertex in which
momentum is going to be conserved. First we must slightly modify the
free particle action (\ref{3}). By choosing appropriately the form
of the affine parameter $\tau$, one can restrict its range to be
from $-\infty$ to $0$ for the incoming particles and from $0$ to
$\infty$ for the outgoing ones. I will implicitly assume this choice
and not write the range of integration in what follows. Thus the
``bulk'' free particles action reads
\begin{equation}\label{5}
    S_{free}=-\sum_{{\cal I}}\int d\tau \,x_{\cal I}^a\dot p^{\cal I}_a
    +N_{\cal I}\left(\eta^{ab}p^{\cal I}_ap^{\cal I}_b-m_{\cal I}^2\right)
\end{equation}
In order to include interaction one has to add the momentum
conservation at the vertex. This condition is imposed by the adding
to the action (\ref{5}) the following  term
\begin{equation}\label{6}
  S_{int}= z^a \widetilde{\sum_{{\cal I}}} p^{\cal I}_a\,,
\end{equation}
where $z^a$ is the Lagrange multiplier imposing the momentum
conservation and the tilde over the sum indicates that the incoming
(four-) momenta are to be taken with the plus, while the outgoing
with the minus, sign. The bulk equations of motion resulting from
the action $S_{free}+S_{int}$ are just the standard free particle
equations (\ref{3a}), for each particle. The equation of motion for
the variable $z^a$ is the momentum conservation at the vertex. The
only remaining equation follows from the boundary contribution at
$0$ to the variation over $p^{\cal I}_a$ combined with variation of
the $S_{int}$ over this variable, which leads to the condition
\begin{equation}\label{7}
x_{\cal I}^a(0) = z^a\,, \forall_{\cal I}\,.
\end{equation}
The meaning of this last condition is clear: the worldlines of all
the particles meet at the interaction point, in which the physical
coordinates of the particles are all equal the ``interaction
coordinate'' $z^a$. Therefore the interaction is local in spacetime,
and  all inertial observers (Lorentz transformed, translated) agree
that the interaction is local. Locality is {\em absolute}.

Let us now investigate what kind of the implicit geometrical data
concerning the flat momentum space have been used to write the
interaction vertex (\ref{6}). Clearly, in order to do so we had to
know how to add momenta and what is the negative momentum. In the
case of momentum space being a linear space both these operations
are easy to define: For the summation we take vectors from the
origin to the points ${\cal P}^{\cal I}$ with coordinates $p^{\cal
I}_a$ and simply sum them. The negative momentum can be constructed
by taking such a vector, multiply it by $-1$ and find which point of
the manifold the tip of the resulting vector points to. Clearly this
is going to be the point $-{\cal P}^{\cal I}$ with coordinates
$-p^{\cal I}_a$. It is worth stressing the obvious fact that the
interaction vertex is invariant under permutation of incoming and/or
outgoing particles, since the momentum conservation rule is abelian.

We will generalize now the flat momentum space particle mechanics to
the case of curved momentum space, pointing out the differences
between these two cases and the new effects that arises as a
consequence of the momentum space curvature. In what follows I will
restrict myself mainly to the case when the momentum space has the
structure of a group manifold. This is motivated by simplicity of
such a choice and also by the $2+1$ dimensional experience. More
general cases can be also considered (see
\cite{AmelinoCamelia:2011bm} and \cite{Freidel:2011mt}).

Let us start with the generalization of the kinetic term in the
action (\ref{5}). Since we have to do with the curved momentum space
manifold now, the coordinate of a point ${\cal P}$ is now $p_\mu$
with ``curved'' index $\mu$. Therefore we must use a tetrad of the
curved momentum space metric $e_a{}^\mu$ (notice the upside-down
position of indices, as compared with the curved spacetime case) and
the natural generalization of the kinetic term reads
$$
- x^a \,e_a{}^\mu(p) \, \dot p_\mu\,.
$$
It should be stressed that since $e_a{}^\mu(p)$ is a nonlinear
function of momenta one needs to have a scale of mass available to
define it
$$
e_a{}^\mu(p) = \delta_a^\mu + \frac{1}{M_P} C^{\mu\nu}{}_a\, p_\nu +
\ldots
$$
Even in the case of a Lie group momentum manifold there are many
metrics that one can use to form the tetrad $e_a{}^\mu$ (see e.g.,
\cite{Milnor2} for clear discussion.) However in this case there is
a natural construction of the kinetic term. To see how it works
notice that on any Lie group there is a natural Lie algebra valued
one form
$$
\omega = g^{-1}dg\,,\quad g \in G\,.
$$
Since this form is Lie algebra valued, we can always write it in
some coordinates as $g^{-1}dg=e_a{}^\mu(p)\, dp_\mu\, \gamma^a$,
where $\gamma^a$ is some basis of the Lie algebra. Defining now the
basis of the dual space by
$$
\left< \sigma_a, \gamma^b\right> = \delta^b_a\,,
$$
we see that the kinetic term can be compactly rewritten as
\begin{equation}\label{8}
    -\left< x, g^{-1}\dot g\right>
\end{equation}
with $x\equiv x^a\sigma_a$ is an element of a space dual to the Lie
algebra of the Lie group $G$.

Having discussed the kinetic term we can turn now to the mass shell
condition which would generalize the standard $p^2-m^2$ of special
relativity. The generalization is  rather obvious once one realizes
that $p^2$ has a geometric meaning of a square of the flat momentum
space distance between the point $\emptyset$ with coordinates
$p_a=0$ (i.e., zero energy and linear momentum) and the point ${\cal
P}$ with coordinates $p_\mu$. Therefore, in the curved momentum
space case one could do exactly the same, defining the distance
$D^2(p)$ which is calculated with the help of the metric
$g_{\mu\nu}=\eta_{ab}e_a{}^\mu e_b{}^\nu$ (although, in principle
one can use any other available metric on the group momentum
manifold.) Putting all this together we see that in the case of the
group momentum manifold one has to replace the action (\ref{5}) with
\begin{equation}\label{9}
    S_{free}=-\sum_{\cal I}\int d\tau \left<x, g_{\cal I}^{-1}\dot
    g_{\cal I}\right>
    +N_{\cal I}\left(D^2(p_{\cal I})-m_{\cal I}^2\right)\,.
\end{equation}

With the free particles action (\ref{9}) at hands, one can now turn
to generalizing the interaction term (\ref{6}). Since in our case
the momentum space  is a group manifold, we can make use of the
group structure to define the momentum composition rule. (In the
more general case one has to introduce the composition structure. It
is shown in \cite{AmelinoCamelia:2011bm} and \cite{Freidel:2011mt}
that this amounts to defining a connection on the momentum space.
Such connection does not need to be, in general, a Levi-Civita
connection of the metric of the tetrad in the kinetic term and/or
the one implicitly present in the mass-shell condition ).

With every point on the momentum group manifold ${\cal P}$ we can
associate a group element $g({\cal P})$. Let us define a function
$K: G\rightarrow R^4$ such that for the group element $g({\cal P})$,
$K_\mu(g({\cal P}))$ are coordinates of the point ${\cal P}$.
Suppose now that we have a vertex with two incoming momenta
$p_\mu^{(1)}$ and $p_\mu^{(2)}$, and the outgoing one $p_\mu^{(3)}$.
Making use of the group structure one defines the momentum
conservation at the vertex by
\begin{equation}\label{10}
    g^{-1}_{(1)}g^{-1}_{(2)}g_{(3)} =1 \mbox{ or }
    K_\mu\left(g^{-1}_{(1)}g^{-1}_{(2)}g_{(3)}\right) = 0
\end{equation}
(notice that with the incoming momenta I associated inverse group
elements in agreement with the convention used in (\ref{6}).) It is
important to realize at this point that the group multiplication is
not abelian in general and thus the ordering of group elements in
(\ref{10}) matters. In the case of trivalent vertex with two
incoming and one outgoing particles there are two possible momentum
conservation conditions
$$
K_\mu\left(g^{-1}_{(1)}g^{-1}_{(2)}g_{(3)}\right) = 0 \mbox{ and }
K_\mu\left(g^{-1}_{(2)}g^{-1}_{(1)}g_{(3)}\right) = 0\,.
$$
We will not dwell here into the discussion which of the two (and
many more for higher valent vertices) has to be chosen and how this
can be decided; we will just assume that each vertex comes with a
priori decided ordering of incoming and outgoing lines. Then the
momentum conservation rule (\ref{6}) is replaced by the analogous,
nonlinear expression
\begin{equation}\label{11}
    z^\mu\, K_\mu\left(p^{(1)}, p^{(2)}, p^{(3)}\right)\,.
\end{equation}
where
$$
K_\mu\left(p^{(1)}, p^{(2)},
p^{(3)}\right)=K_\mu\left(g^{-1}_{(1)}(p^{(1)})g^{-1}_{(2)}(p^{(2)})g_{(3)}(p^{(3)})\right)\,,
$$
if the first conservation rule above is used.

 The bulk  equations of
motion following from the action (\ref{9}) are (for each particle,
so that we do not add the label ${\cal I}$)
\begin{align}
&e^\mu_a\, \dot p_\mu =0 \label{12}\\
&\frac{d}{d\tau}\left(e^\mu_a\, x^a\right) -  \frac{\partial
D^2}{\partial p_\mu}=0\label{13}\\
&D^2(p)-m^2=0\label{14}
\end{align}
where in (\ref{14}) we make use of the reparametrization of $\tau$
invariance to set $N=1$. Notice that it follows from these equations
that the momentum is conserved along the wordlines, $\dot p_\mu=0$,
because the tetrad $e^\mu_a$ in eq.\ (\ref{12}) is invertible by
definition.

The equations (\ref{12})--(\ref{14}) should be appended by equations
resulting from the presence of the vertex. One of them comes from
variation over $z^\mu$ and enforces the momentum conservation at the
vertex
\begin{equation}\label{15}
K_\mu\left(p^{(1)}, p^{(2)}, p^{(3)}\right)=0\,,
\end{equation}

 The second comes from
the boundary contribution at $\tau=0$ of the variations over momenta
of the bulk particle actions (we will ignore a possible contribution
of the boundaries at $\tau=\mp\infty$ for incoming/outgoing
particles), to wit
\begin{equation}\label{16}
   \bar x_{\cal I}^a\, e^a_{\mu}(p^{\cal I}) = z^\nu\, \frac{\partial
    K_\nu}{\partial p^{\cal I}_\mu}\,,\quad  \bar x_{\cal I}^a\equiv x_{\cal I}^a(0)\,.
\end{equation}
This condition relates the coordinates in two spacetimes: the
ambient spacetime with coordinates $x^a$ and the spacetime of
``interaction coordinates'' $z^\mu$ which can be associated with the
(co)-tangent space to the momentum manifold at the origin. It is
important to notice that this relation depends in general on all the
momenta carried by the particles whose worldlines meet at the
vertex. This momentum dependence disappear only in the case in which
$K_\mu$ is a linear function of momenta and $e^a_{\mu}=\delta^a_\mu$
i.e., if and only if the momentum space is flat.

Before turning into the discussion of physical interpretation and
consequences of this result, let us return to the symmetries of the
action (\ref{9}), (\ref{11}). we will discuss here only the
translational symmetry (the delicate issue of Lorentz symmetry is
addressed in \cite{Gubitosi:2011ej}.) The kinetic term of (\ref{9})
is invariant, up to a boundary term, under the following global
translational symmetry, with parameter $\xi$, generalizing
(\ref{4}). To see how it works let us first rewrite the kinetic term
as
$$
-x^\mu_{\cal I}\, \dot p^{\cal I}_\mu\,, \quad x^\mu_{\cal I}\equiv
x^a_{\cal I}\, e^\mu_a(p^{\cal I})\,.
$$
The most general transformation that leaves the bulk action
$$
 S_{free}=-\sum_{\cal I}\int d\tau x^\mu_{\cal I}\, \dot p^{\cal I}_\mu
    +N_{\cal I}\left(D^2(p_{\cal I})-m_{\cal I}^2\right)
    $$
invariant up to a boundary term reads
$$
\delta x^\mu_{\cal I} = \xi^\nu\frac{\partial F_\nu(p)}{\partial
p^{\cal I}_\mu}
$$
The resulting boundary term in variation of the action is $-\xi^\nu
\dot F_\nu$ and it can be canceled by the vertex term if $F_\nu =
K_\nu$ and $\delta z^\nu = \xi^\nu$ and therefore the total action
is invariant under the translations
\begin{equation}\label{17}
    \delta x^\mu_{\cal I} = \xi^\nu\frac{\partial K_\nu(p)}{\partial
p^{\cal I}_\mu}\,,\quad \delta z^\mu=\xi^\mu\,,\quad \delta p=\delta
N =0\,.
\end{equation}
Returning to the original variables we find
\begin{equation}\label{18}
    \delta x^a_{\cal I} = \xi^\nu\frac{\partial K_\nu(p)}{\partial
p^{\cal I}_\mu}\, e_\mu^a(p^{\cal I})\,.
\end{equation}
It should be noticed that an extension of this result to the case of
many vertices is not straightforward, see
\cite{AmelinoCamelia:2011nt} for the detailed discussion.

Equation (\ref{16}) along with the equations (\ref{17}) and
(\ref{18}) governing the translational invariance lead us directly
to the issue of relative locality. As a result of the translational
invariance (\ref{17}) there exist an observer for whom the
interaction coordinates of the process $z^\mu$ are zero. Then it
follows from (\ref{16}) that the particles coordinates $\bar
x^a_{\cal I}$ for all the particles vanish as well. For this
particular observer the process is {\em local}. However {\em
locality is relative}: for any translated observer $\bar x^a_{\cal
I}\neq \bar x^a_{\cal J}$ if ${\cal I}\neq {\cal J}$ the
translations of the wordlines in the ambient space with coordinates
$x^a_{\cal I}$ depend not only on the momentum carried by the
particle ${\cal I}$ but also on the momenta carried by all the
particles interacting in the vertex. This is a striking novel
feature of the curved momentum space as compared to special
relativity where locality has an absolute meaning, as all the
special relativistic observers agree on what is local and what is
not. The principle of relative locality is discussed further in
\cite{AmelinoCamelia:2011bm} and \cite{AmelinoCamelia:2011pe}.

This completes this brief discussion of the theory of particles with
curved momentum space. There are many open problems that must be
still investigated. The most pressing one concerns the spacetime. As
we showed above there are two spacetimes involved in the
construction: the ambient spacetime with $x^a$ coordinates and the
interaction spacetime with coordinates $z^\mu$, which coincide in
special relativity and are approximately identical if the
energies/momenta of particles involved in the process are very small
as compared with the momentum scale $M_P$. The question arises as to
which of them is physical i.e., in which the operatively well
defined spacetime measurements (of distance, speed, etc) are taking
place. This is an important question because it is likely that
theories with curved momentum space might be testable by this type
of measurements (see \cite{Freidel:2011mt} and
\cite{AmelinoCamelia:2011nt}.)

Another problem is to turn the qualitative argument leading to
curved momentum space in four dimensions, presented at the beginning
of this article to a solid derivation from the first principles.
Work on this question is in progress.

Last but not least it would be of great interest to extend the
construction from particles to fields. At this moment only some
particular models of free scalar field with curved momentum space
are known, and it is not known how to generalize them to interacting
theories involving higher spin fields (especially spinors and
Yang-Mills fields.) It would be very important to be able to
construct at least the leading order corrections to the Standard
Model lagrangian (as it has been done in the case of theories with
Lorentz invariance violations.) If successful this research program
will make it possible to investigate the consequences of curved
momentum space and relative locality and make contact with
elementary particles phenomenology.

\section*{Acknowledgment} This work was supported in parts by the
grant  2011/01/B/ST2/03354 and by funds provided by the National
Science Centre under the agreement DEC-2011/02/A/ST2/00294.

\end{document}